\documentclass[review]{acta}
\usepackage{epsfig}
\usepackage{amsmath , amssymb , latexsym}
\usepackage{epstopdf}
\usepackage{tabularx}
\usepackage{float}
\usepackage{siunitx}
\usepackage{graphicx}
\usepackage{hyperref}
\hypersetup{
    colorlinks=true,
    linkcolor=blue,
    filecolor=magenta,
    urlcolor=cyan,
}

\SetPages{369}{380}

\SetVol{69}{2019}

\begin{document}

\begin{Titlepage}

\Title{Central wavelengths and profile shapes of diffuse interstellar bands vs. physical parameters of intervening clouds
\footnote{This paper includes data gathered with the
8~m telescope at the Paranal Observatory and with the 6.5 Magellan telescope at the Las Campanas Observatory (Chile).}}

\Author{K~r~e~{\l}~o~w~s~k~i,~J., M~a~r~i~{\'c}, ~T., K~a~r~i~p~i~s, ~A.~~and~~S~t~r~o~b~e~l,~A.,}{Toru\'n Centre for Astronomy,
Nicolaus Copernicus University, Gagarina 11, 87-100 Toru{\'n}, Poland \\
e-mail: jacek@umk.pl; tanj.maric@gmail.com; andrew.karipis@gmail.com; strobel@umk.pl}
\Author{G~a~l~a~z~u~t~d~i~n~o~v, ~G.~A.}{Instituto de Astronomia, Universidad Catolica del Norte, Av. Angamos 0610, Antofagasta, Chile\\
       Pulkovo Observatory, Pulkovskoe Shosse 65, Saint-Petersburg, Russia\\
       e-mail: runizag@gmail.com}

\Received{\today}
\end{Titlepage}

\Abstract{This paper tries to establish whether there are variations of the central wavelengths or the profile shapes of
diffuse interstellar bands (DIBs) and whether these variations are caused by different physical parameters of translucent clouds.
 For this purpose we used spectra of two stars seen through two different single clouds: HD34078 (AE Aur) \& HD73882 acquired using
 two different instruments:  the MIKE spectrograph, fed with the 6.5 m Magellan telescope at Las Campanas Observatory, and the UVES,
fed with the 8 m Kueyen telescope at the Paranal observatory. The wavelength displacements of the DIBs at 6196, 6203, 6376, 6379 and 6614\si{\angstrom}
with respect to the well known interstellar atomic and molecular
lines (K{\sc i} and CH) have been measured. The mentioned shift is seemingly absent in the DIBs at 4726, 4964, 4763, and 4780\si{\angstrom}.
In addition the considered profiles may show (in HD34078) extended red wings. The observed phenomena are likely related to physical parameters of
intervening clouds (rotational temperatures of molecular species) and may help in the identification of the DIB carriers.}
{Galaxy : disk; (ISM:) dust, extinction; (ISM) lines, bands.}

\section{Introduction}

Translucent interstellar clouds exhibit several phenomena being revealed by interstellar absorption features. The most prominent one is \textit{extinction} of
starlight caused by interstellar dust. Absorption spectra  of translucent clouds also contain well-known \textit{atomic lines} and \textit{molecular bands} of
simple polar radicals (usually two atom radicals: CH, CN, CH$^+$). Furthermore, rotational emission features revealed  the presence of many complex molecules
(in star-forming regions), the largest currently known are the fullerenes C$_{60}$ and C$_{70}$
(Cami et al. 2010; Berne et al. 2013). This clearly displays rich, carbon-based chemistry in interstellar clouds. The feature that makes up the longest standing
unsolved question in astronomical spectroscopy is conveyed by \textit{diffuse interstellar bands (DIBs)} - for a recent review see e.g. Kre{\l}owski (2018).

First diffuse interstellar bands have been discovered by Heger in 1922 (Heger 1922). The application of state of the art detectors and telescopes to observations
led to the progressive discovery of many additional weak features, almost all of them in the range from the near–UV to the near–IR. The current lists of known
DIBs contain over 400 entries (Hobbs et al. 2009; Galazutdinov et al. 2000).
All conceivable forms of matter have already been proposed as DIB carriers: from interstellar hydrogen negative ion to dust grains. The
most common view is that these molecules are some form of polycyclic aromatic hydrocarbons (PAH’s), carbon chains or fulleranes.  Unfortunately all
specific proposals have been rejected or disputed, and after almost a century DIBs carriers remain unidentified.  Despite many laboratory-based studies of potential
carriers, it has not been possible to unambiguously link these bands to specific molecular species. This is unfortunate, as  such the link would
significantly contribute to our understanding of evolution of interstellar medium.

In most stars we do observe similar wavelengths of DIBs in relation
to the atomic and molecular features in the same spectra. However,
DIBs' wavelengths are not constant and some physical conditions may
change their rest wavelengths, causing either blue or red shifts.
This effect was first described by Kre{\l}owski \& Greenberg (1999)
-- red-shift in Ori OB1 objects. Galazutdinov, Manic{\'o} \&
Kre{\l}owski (2006) suggested a blue-shift of certain DIBs in the
spectrum of AE Aurigae (HD34078). A similar blue-shift have been
discovered also in the Sco OB1 association (Galazutdinov et al.
2008) but it's cause remains uncertain. The uncertainty follows the
fact that some Doppler shifted components, even if weak, are
observed in atomic/molecular features. It is interesting whether
all DIBs suffer the same shift if being observed in the same
object. Galazutdinov et al. (2013) reported a detection of
variability in interstellar absorption lines of Ca{\sc i} at
4227~\AA\ and Fe{\sc i} at 3860~\AA\ in spectra of HD73882 acquired
in 2006 January and 2012 January. Intensity and profile shapes of
the Ca{\sc i} and Fe{\sc i} lines had dramatically changed within
the 6 year period. Other interstellar features, observed along the
same line of sight, do not demonstrate strong changes. It is likely
that a high velocity CaFe cloud (Bondar et al., 2007) obscured the
star between the two observations.

A vast majority of reddened stars is observed through several
interstellar clouds. As a result the detected DIBs are ill-–defined
averages which are unfit for discerning physical parameters and
spectra of individual clouds. The fact that DIBs are most
frequently observed through a number of possibly very different
clouds is likely to be one of the reasons for their apparent
uniformity, washing out individual differences. This encourages us
to look for extraordinary spectra, most likely formed in single
clouds, and extract a more sensible physical interpretation. Some
of the known, reddened stars, apparently shine through rather
specific or ``peculiar'' clouds. One of them is HD34078 where the
rotational temperature was estimated as especially high by
Adamkovi{\'c} et al. (2003). Diffuse band's profiles were suggested
as peculiar (Galazutdinov, Manic{\'o} \& Kre{\l}owski 2006). Some
of the molecular features, seen in its spectrum show relatively
rapid changes with time which may suggest that the intervening
cloud is situated closely to the star (Kre{\l}owski et al., 2016) and
is influenced by its radiation.

\section{Observational material}

We observed two objects, HD 73882 and HD 34078 (AE Aurigae) that
are heavily reddened objects free of evident Doppler splitting in
atomic or molecular lines. The latter property indicates that these
stars may represent homogeneous but different environments; their
comparison may shed some light on our problem. The selected targets
seem to be relatively close; however their distances appear still
uncertain. Their trigonometric parallaxes have recently been
measured by Gaia. However, we decided to compare the Gaia distances
to other possible estimates. Ca{\sc ii} lines apparently can
provide information related to the distance of early-type stars
additively to trigonometric parallaxes (Megier et al., 2009). Our
obtained Ca{\sc ii} distances (for our two targets) are similar to
those calculated using photometric equation (from Sp/L). However,
the Gaia data lead to much shorter distances in both cases. These
distances together with other parameters of our stars are given in
Table 1. \linebreak

\begin{table}[H]
\label{stardata}
\resizebox{\textwidth}{!}{%
\begin{tabular}{|ccccccccccccc|}
\hline
Star  & Sp/L    & V    & B-V  & D(Sp/L) & $\pi$(Gaia) & $\pm$ & D(Gaia) & EW(K) & $\pm$ & EW(H)    & $\pm$ & D(CaII) \\ \hline
73882 & 08.15IV & 7.19 & 0.40 & 1175    & 2.89        & 0.45  & 350     & 281.8 & 1.6   & 1172.6   & 1.2   & 1200    \\
34078 & 09.5V   & 5.96 & 0.22 & 535     & 2.46        & 0.07  & 405     & 130.2 & 1.7   & 79.5     & 1.6   & 590     \\
\hline
\end{tabular}%
}
\caption{Comparison of three method distance measurements.}
\end{table}

If to apply the Gaia parallax to HD73882 the total--to--selective extinction ratio should be as large as 6.9 which clearly contradicts the value given by Fitzpatrick \& Massa (2007) --
which is almost exactly equal to the galactic average. In the case of HD34078 the ``Gaia R$_V$'' is about 4.4 also evidently higher than that estimated by Fitzpatrick \& Massa.

 Establishing the rest wavelengths of DIBs is not a trivial task as they have never been observed in laboratories. Our approach is based on the assumption that shifting the spectra of
reddened stars to the rest-wavelength frame of either  methylidyne (CH) or neutral potassium (K{\sc i}) lines shifts also DIBs to their laboratory wavelengths. The diffuse bands which
can be seen in our stars' spectra and whose behaviour we will discuss afterwards are at about : 6196, 6203, 6376, 6379, 6614, 4964, 4763, 4780, and 4726 \si{\angstrom}. We do not
discuss the major 5780 and 5797 DIBs as the latter are out of the range covered with UVES spectrograph.
\newline
An important confirmation of the reality of any observed phenomena is a comparison of data from at least two instruments,
which is especially important while searching for very subtle effects. The spectra were recorded with the aid of the  Magellan/Clay telescope at the Las Campanas
Observatory in Chile using the Magellan Inamori Kyocera Echelle (MIKE) spectrograph (during the same night)  and the Ultraviolet and Visual Echelle Spectrograph (UVES)
at Paranal Observatory also in Chile.

All of the data were reduced using DECH data reduction suite.

General information about the observed spectra is given in Table 2.

\begin{table}[H]
\centering
\label{listspectra}
\begin{tabular}{|c|c|c|c|c|}
\hline
Star    & Instrument  & Resol.   & Date        & SNR  \\
\hline
34078   & UVES        & 80,000   & 2005-12-18  & 300  \\
34078   & MIKE        & 83,000   & 2012-01-12  & 680  \\
73882   & UVES        & 80,000   & 2003-11-13  & 170  \\
73882   & MIKE        & 65,000   & 2012-01-11  & 800  \\
73882   & UVES        & 80,000   & 2014-03-12  & 760  \\
\hline
\end{tabular}
\caption{List of spectra of both objects}
\end{table}

We used two UVES spectra of HD73882 because one of them covers only the blue range while the other is of much lower S/N ratio.

\section{Method}

Prior to the measurements of displacement of diffuse bands, the wavelength scale of all spectra was shifted to move the CH $ \lambda$4300 \si{\angstrom} and and K{\sc i} $ \lambda$7699 \si{\angstrom}
lines to their laboratory positions. The selection of CH as a rest point is argued for by the fact that in most surveys the radial velocities of the cores of profiles of CH $ \lambda$4300 \si{\angstrom}
and K{\sc i} $ \lambda$7699 \si{\angstrom} are almost identical (Galazutdinov et al. 2000).
Our analysis may allow us to divide the observed DIBs into several sets using this new criterion: the behaviour of profiles and  positions of central wavelengths.

\begin{center}
\begin{figure}[H]
\label{fig1}
\includegraphics[scale=0.45]{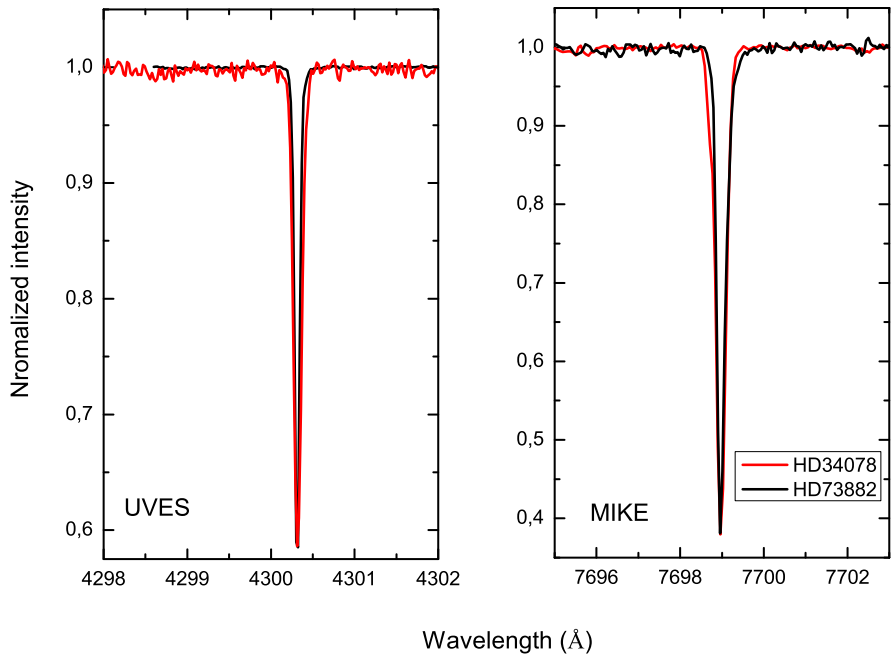}
\caption{Left: the CH line from UVES. Right: the K{\sc i}  line from MIKE. The lines are normalized to the same depth.}
\end{figure}
\end{center}

wavelength scale using K{\sc i} line at 7699 \si{\angstrom} and CH
line at 4300 \si{\angstrom}. It is important to make sure that the
observed effects (in DIBs) are not simply Doppler shifts. The
latter should be observed in atomic or molecular lines as well and
is thus disproved beyond a doubt by its lack. Positions of the
lines carried by simple interstellar radicals coincide in both
targets which suggests that all possible displacements of
interstellar features are likely caused by physical/chemical
conditions in the clouds situated along the sight lines to our
targets.

We used unreddened stars Regulus and Spica as telluric lines divisors for our spectra of HD73882 and HD34078. Part of spectra containing DIBs at 6270 and 6284 \si{\angstrom} from UVES is heavily
contaminated by telluric features and extraction of any viable information was inconceivable. HD34078 is very low at the Southern Sky. The largest uncertainty regarding these measurements stems
from establishing the continuum, especially if we are dealing with broad features. We filtered out white noise using Fourier 5 filter from DECH suite.

\section{Results}
Every calculated shift of observed interstellar bands is confirmed by two instruments and given in Table 3.

\begin{table}[tbp]
\centering

\label{wshift}
\begin{tabular}{|c|c|c|}
\hline
 DIB wavelenght (\si{\angstrom})
     &  MIKE (km/s) & UVES (km/s)   \\
 \hline
4726 & no shift           &    no shift       \\
4763 & no shift           &    no shift      \\
4780 & no shift           &    no shift      \\
4964 & no shift           &    no shift       \\

6196 &-10.9$\pm$1.3    & -9.4$\pm$0.6  \\
6203 & -8.6$\pm$2.2    & -8.6$\pm$0.5  \\
6376 & -6.8$\pm$1.7    & -5.7$\pm$1.1  \\
6379 & -8.8$\pm$1.1    & -9.5$\pm$1.2  \\

\hline
\end{tabular}
\caption{Measured shift of diffuse bands in spectra of AE Aur in respect to position of those in spectra of HD 73882  }
\end{table}

\begin{center}
\begin{figure}[H]
\label{fig1}
\includegraphics[scale=0.45]{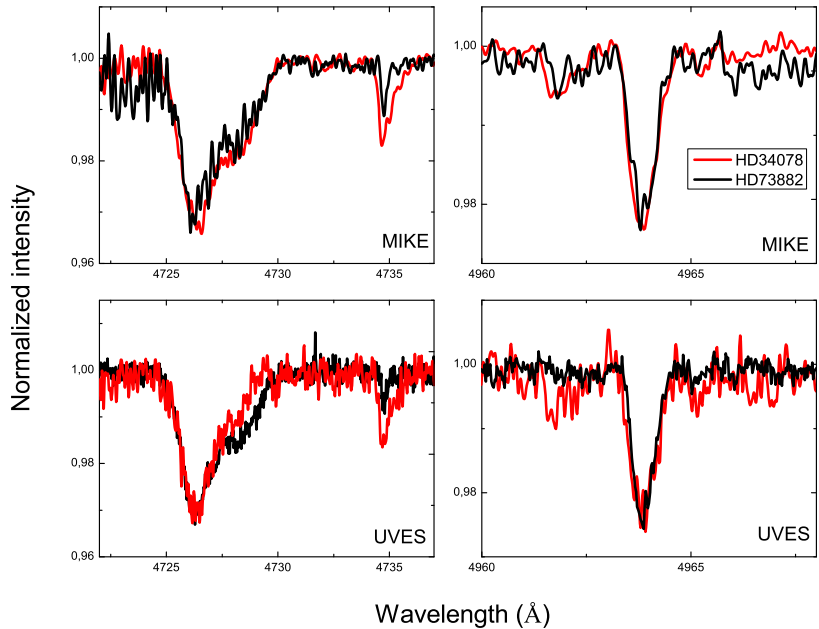}
\caption{Left: the 4726 DIB. In the upper panel spectra from MIKE and in the lower from UVES are presented. Right: the 4964 DIB. In the upper panel spectra from MIKE and in the lower from UVES. No shift is observed.}
\end{figure}
\end{center}


Fig. 2 depicts the 4726 and 4964 DIBs which are evidently not blue--shifted in the spectra of AE Aurigae when compared to the reference star HD 73882.
It seems clear that the 4726 DIB  shows no shift in the core but may display an extended red wing.

The point to be emphasized is that the rotational temperature of AE
Aurigae of the $C_{3}$ molecule is very high (Adamkovic et al.
2003). That in the direction of HD73882 should be very low as the
$C_{3}$ band, seen in its spectrum, closely resemble those of
HD169454 or HD204827 mentioned as very low temperature cases by
Adamkovi{\'c} et al. (2003).

\begin{figure}[H]
\centering
\label{fig2}
\includegraphics[scale=0.45]{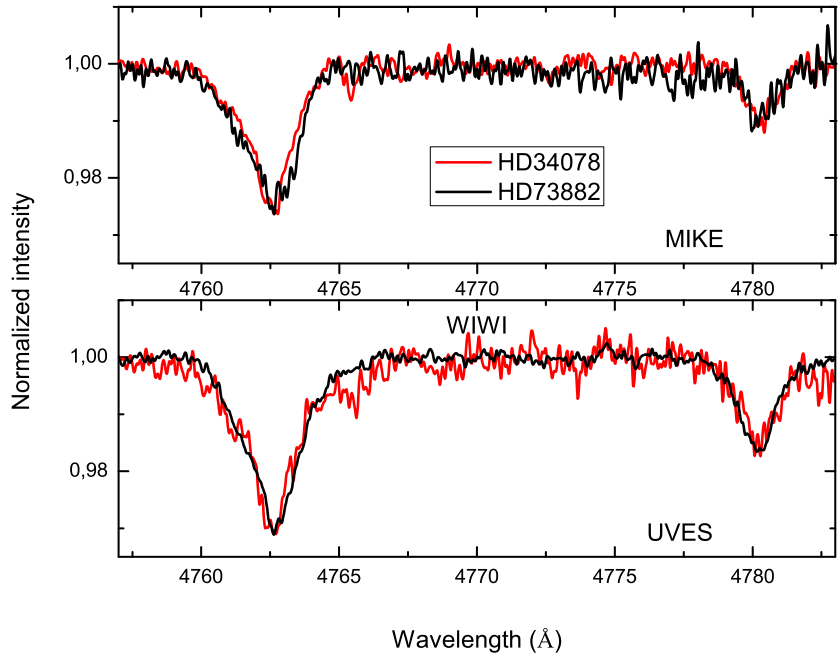}
\caption{ The 4763 \& 4780 DIBs. The upper panel depicts MIKE spectra while the lower -- UVES ones. The central depths are artificially made identical to allow profile comparison. }
\end{figure}


Figure 3 portrays the DIBs at 4763 and 4780 which are apparently
not shifted in the spectra of both stars. Their profiles are of the
same strength ratio and the same shapes i.e. do not seem to be
related to rotational temperatures of molecular species. Possibly
they are produced by the same carrier.

\begin{figure}[H]
\centering
\label{fig4}
\includegraphics[scale=0.45]{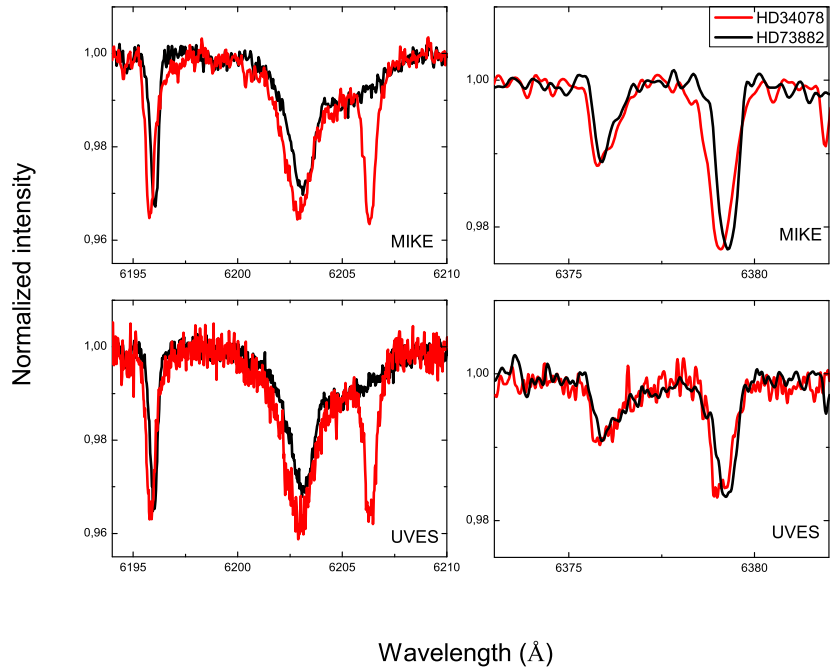}
\caption{Left: the 6196 \& 6203 DIBs. In the upper panel from MIKE and in the lower -- from UVES. Right panels: the 6376 \& 6379 DIBs. The upper panel -- spectra from MIKE, the lower one -- from UVES. The wavelength shift is evident in all DIBs and in both instruments.}
\end{figure}

\vspace{-1em}
\bigskip

Figure 4 shows evident blueshits of DIBs at  6196 and 6203 which, only in AE Aur, may be accompanied with some red wings. In AE Aurigae spectra there is a stellar line but it does not contaminate the nearby band (6203). Also the DIBs at 6376, and 6379 seem to be blueshifted in the spectra of HD34078 (AE Aur) while compared to HD73882.

\begin{figure}[H]
\centering
\label{fig5}
\includegraphics[scale=0.45]{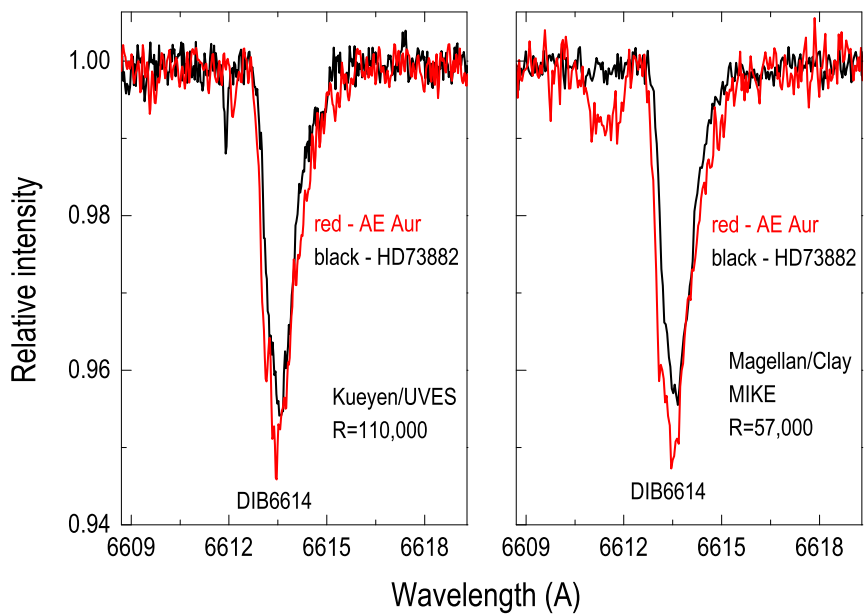}
\caption{Left: the 6614 DIB seen using both instruments. Left panel spectra from UVES. Right panel: spectra from MIKE. The wavelength shift is not very evident but the profile is ``blue--broadened'' in AE Aur and shows the extended red wing. }
\end{figure}

\vspace{-1em}
\bigskip

The last strong DIB, being available to both our instruments is the 6614 one. Its profile in both our objects is depicted in Fig. 5. It is evident that the profile in HD34078 is broader --
like if adding some blue component and has an extended red wing. The possible wavelength shift is not very evident.

From our analysis, we see that the phenomena of wavelength shifts are not randomly distributed over the wavelength range, but quite concentrated in yellow-red segment of spectrum.
It is important to note that some DIBs do not share the blue-shifts which indicates that they are carried by different species.

We should also remind that we skipped the major 5780 band 5797 DIBs
because they are not covered by the UVES spectra. However, the 5797
DIB profile seems to show the same shift as 6614 if only MIKE
spectra are compared.

\section{Discussion}

The most probable explanation for this shift is the rotational
temperature of the carriers. The temperature in interstellar cloud
is an intensive parameter, but the equivalent width of DIB is an
extensive one. It may be argued for (Adamkovi{\'c} et al. 2003)
that the profile shapes of certain DIBs correlate with the
populations of excitation levels of interstellar $ C_{2}, C_{3},
CN$. The broadening of certain DIBs with increasing $ C_{2}$
rotational temperature was reported by Ka{\'z}mierczak et al.,
(2009), while the correlation with $T_{01}$ temperature of H$_2$
was mentioned in Gnaci{\'n}ski et al., (2009).
 Another molecule whose rotational transitions correlate with behaviour of DIBs may be $CN$  (Kre{\l}owski et al.  (2012)).  CN is a relatively heavy molecule which leads to
 the quick
 saturation of the strongest transition, but we have restricted the plot given in Figure 6, to the transitions from the first rotationally excited level. It is evident that CN's in the spectra of
 HD34078 and HD73882 are of different rotational temperatures. The strength ratio of the R$_0$ and R$_1$ transitions should be close to $\pi$ number (Kre{\l}owski et al.  (2012))
 which is true in HD73882 but certainly not in HD34078 while CN lines are far from saturation in both depicted cases.

\begin{figure}[H]
\centering
\label{fig5}
\includegraphics[scale=0.3]{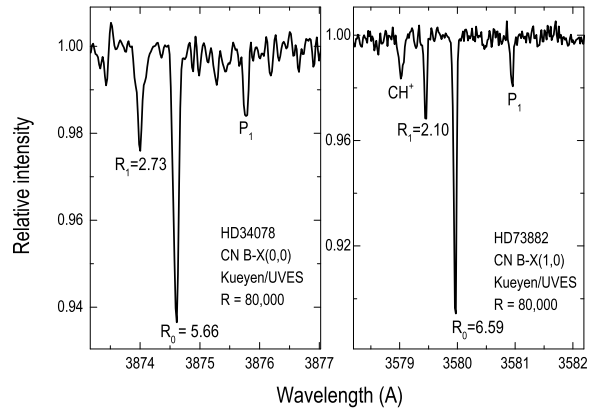}
\caption{Rotational temperature of $CN$ towards our targets. The R$_0$/R$_1$ ratio in HD73882 is 3.14 which suggests the CMBR temperature while that towards HD34078 is 2.07
which must follow a much higher rotational temperature. }
\end{figure}

Populated higher rotational levels of the mentioned molecules toward HD 34078 reflect the presence of high UV radiation flux in clouds located in front of the star which is a
rare and important example of different physical conditions in the interstellar matter. This is illustrated in Figure 7 where we observe the difference in intensity of
higher R number $ C_{2}$  transitions in two objects of interest.

\begin{figure}[H]
\centering
\label{fig6}
\includegraphics[scale=0.3]{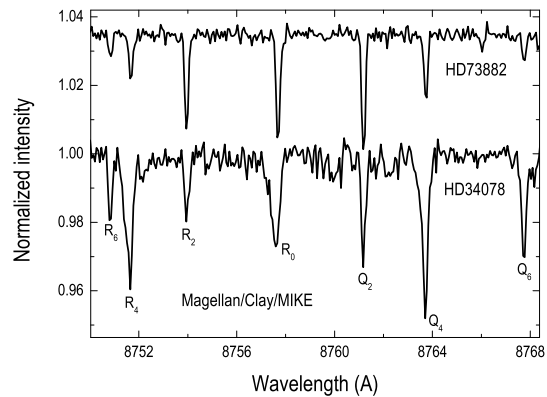}
\caption{Rotational temperature of $C_{2}$. Note the strong transitions from high rotationally excited levels in AE Aur which indicate for much higher rotational temperature.}
\end{figure}

 Furthermore, linear carbon chain as well as any centrosymmetric molecule has no permanent dipole moment and thus lacks the ability to radiate excess energy away.
 Therefore, rotational temperature of these molecules is higher than the one expected in interstellar medium (that of cosmic background radiation). We believe that
 analogous to variation in profile of these bands, variation of profiles of diffuse interstellar bands is caused by growing population of higher rotational levels in
 the DIB carriers. It is clear from Fig. 8 that this change could induce what we observe as shift due to interpreting band head as band center. Let's remind that
 the C$_3$ rotational temperature was estimated as exceptionally high in AE Aur by Adamkovi{\'c} et al. (2003).

\begin{figure}[H]
\centering
\label{fig8}
\includegraphics[scale=0.4]{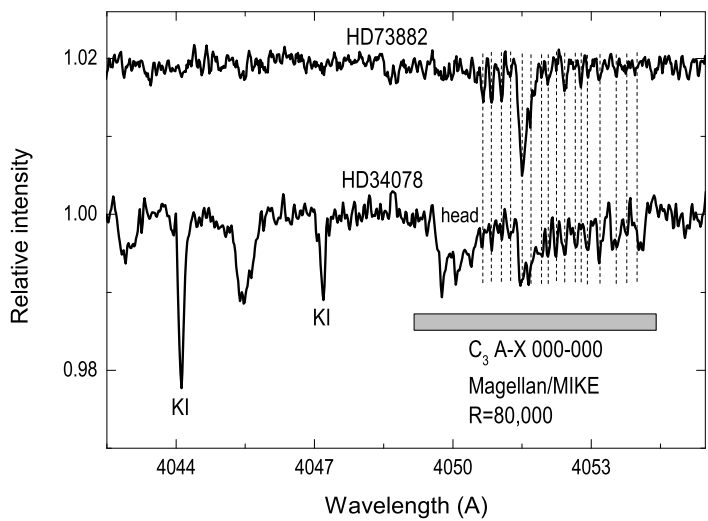}
\caption{Figure demonstrates profile change of $C_{3}$ in the two observed stars with radically different rotational temperature of the mentioned molecule.}
\end{figure}
\vspace{-1em}
\

 It seems that after having carefully analysed our spectra, this blue shift can not be explained in terms of a simple phenomena like a Doppler shift.
It is something that has to do with the physical parameters of the
cloud or the difference (between the two targets) of the rotational
temperature between different molecules of the same carrier. The
long term purpose is for us to explain the reason of this shift in
order to acquire a better understanding on the interstellar clouds
evolution.

\section{Conclusions}
The results of this paper can be summed into the following:
\begin{itemize}
\item the rotational temperatures of all molecular species ($CN$, $C_{3}$ and $C_{2}$) are different in spectra of our targets
\item There is evident wavelength blue shift of the DIBs 6196\si{\angstrom},6203\si{\angstrom},6376\si{\angstrom}, 6379\si{\angstrom} and 6614\si{\angstrom}
in the spectra of HD34078 in comparison with those of HD73882
\item These shifts are not a result of Doppler effect, as the CH line at 4300\si{\angstrom} and the potassium line at 7699\si{\angstrom} (showing no Doppler components)
were used to shift the spectra to the rest wavelength velocity frame. Also, it is not instrumental because it is observed in spectra from two different instruments
\item The cause of this shift must lie with the structure of the carrier-molecules of the DIBs. The $C_{3}$ band, shows a strong bandhead (see Fig. 8) and that's what may
be causing the shift. When the maximum energy is absorbed in highly
excited levels, the bandhead shifts towards the blue and a profile
gravity center moves towards the blue also
\item Thus our work confirms that DIB carriers are most likely complex molecules: in simple ones (see C$_2$)  the bands do not resemble the observed DIBs. The rotational
constant is smaller in bigger molecules, meaning that the
subsequent transitions are closer. Figures 6 and 7 show us exactly
that physical parameters of the clouds obscuring our two targets
are radically different
\item Atomic lines may only be moved by the Doppler effect. As for dust grains, it has been proved they don't carry DIBs, because of their profile shapes. As a result,
only molecules can carry such variable features, the observed shifts and deformations most likely caused by physical parameters.

The results of this paper require much discussion and more
research. Especially single cloud cases, observed in very high
resolution and S/N ratio seem to be of basic importance. We only
tried to place small piece to the puzzle that would help to the
identify DIBs carries and solve the longest unsolved spectroscopic
problem: Diffuse Interstellar Bands.
\end{itemize}

\Acknow{JK acknowledges the financial support of the Polish National Science
Center during the period 2015--2018 (grant 2015/17/B/ST9/03397). GAG and JK acknowledge the financial support of the Chilean fund CONICYT REDES 180136.
TM and AK acknowledge support from the Toru{\'n} Astrophysics / Physics Summer program
(TAPS).}

\end{document}